# The Information Catastrophe

Melvin M. Vopson[*]

University of Portsmouth, School of Mathematics and Physics, PO1 3QL Portsmouth, UK

*E-mail: melvin.vopson@port.ac.uk

Currently we produce ~$10^{21}$ digital bits of information annually on Earth. Assuming 20% annual growth rate, we estimate that after ~350 years from now, the number of bits produced will exceed the number of all atoms on Earth, ~$10^{50}$. After ~250 years, the power required to sustain this digital production will exceed $18.5 \times 10^{12}$ Watts, i.e. the total planetary power consumption today, and after ~500 years from now the digital content will account for more than half of the Earth's mass, according to the mass-energy-information equivalence principle. Besides the existing global challenges such as climate, environment, population, food, health, energy and security, our estimates here point to another singularity event for our planet, called the Information Catastrophe.

**1. Introduction**
Since the first discovery of the transistor in 1947 and the integrated microchip in 1956, our society underwent huge technological developments. In just over half a century since the beginning of the Silicon revolution, we have achieved unprecedented computing power, wireless technologies, Internet, artificial intelligence, and multiple technological advances in display technologies, mobile communications, transportation, medicine, to name a few. However, none of these could have been possible without mastering the ability to create and store large amounts of digital information. In fact, digital information is a valuable commodity and the backbone of some of the largest hi-tech companies in the World today. Here we examine the physics of information creation and we determine that, assuming the current growth trends in digital content continue, the World will reach a singularity point in terms of the maximum digital information possibly created and the power needs to sustain it, called the Information Catastrophe.

IBM estimates that the present rate of digital content production is about 2.5 quintillion digital data bytes produced every day on Earth (2.5 × $10^{18}$ bytes or 2.5 billion Gb) [1]. Since 1 byte is made up of 8 bits of digital information, the total number of bits produced on the planet daily is 2 × $10^{19}$. From this we can easily estimate the current annual rate of digital bits production on Earth to be a staggering, $N_b$ = 7.3 × $10^{21}$ bits. Figure 1 shows a chronological list of some of the key technological milestones that enabled the rapid and unstoppable growth of digital information production today.

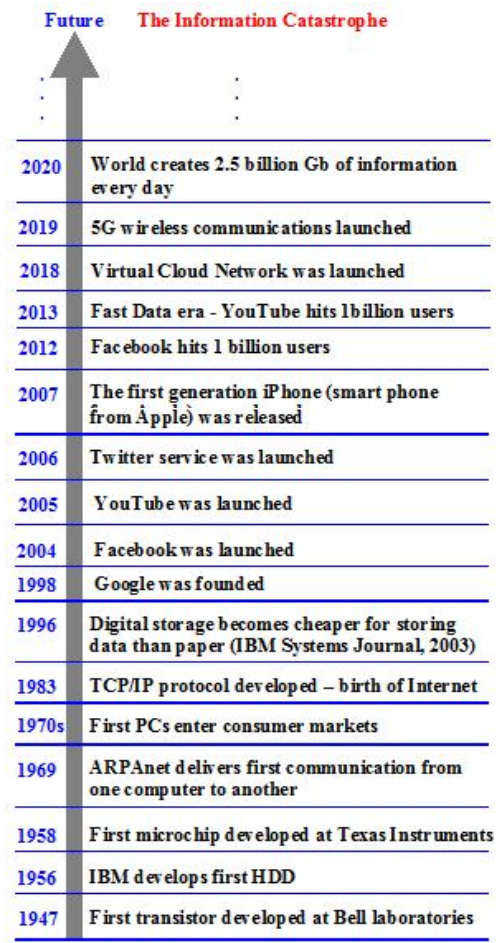

**Figure 1.** Key chronological developments leading to the Information Catastrophe.

## 2. Number of bits estimation

Let us assume that f % is the annual growth factor of digital content creation on Earth. This allows the estimation of the total number of bits of information accumulated on the planet, after n years of f % growth as:

$$N_{bits}(n) = \frac{N_b}{f} \cdot \left( (f+1)^{n+1} - 1 \right) \qquad (1)$$

Although the current estimated f % growth is double-digit, accounting for the fact that out-of-date digital content gets erased all the time, let us assume a conservative annual growth of digital content creation of 1%, i.e. f = 0.01. At this rate, we will have ~$10^{50}$ bits of information after ~6000 years from now. This number of bits is very significant because it represents the approximate number of all atoms on Earth. The size of an atom is ~$10^{-10}$ m, while the linear size of a bit of information today is $25 \times 10^{-9}$ m corresponding about 25 nm$^2$ area per bit, at data storage densities exceeding 1 Tb/in$^2$. Even assuming that future technological progress brings the bit size down to sizes closer to the atom itself, this volume of digital information will take up more than the size of the planet, leading to what we define as The Information Catastrophe.

If we assume more realistic growth rates of 5%, 20% and 50%, the total number of bits created will equal the total number of atoms on Earth after ~1200 years, ~340 years and ~150 years, respectively. It is important to consider that the growth of digital information today is closely linked to other factors including population growth and the increased access to information technologies in the developing countries. If any of these other factors are reversed or saturated, the total number of bits of information accumulated on the planet could display saturation at some point in the future, rather than following equation (1).

**3. Information power estimation**

We now explore these implications from the point of view that a bit of information is not just an abstract mathematical entity, but it is physical. In 1961, Rolf Landauer first proposed a link between thermodynamics and information. Landauer predicted that erasing a bit of information requires a dissipation of energy, equal with at least $k_B T \cdot \ln(2)$, where $k_B$ is the Boltzmann constant and T is the temperature at which the information is stored [2]. Due to the conservation of energy, an energy input of the same value, $k_B T \cdot \ln(2)$, is required to create a bit of information. This is known as the Landauer's principle [3], deduced from thermodynamic considerations and demonstrated experimentally in several recent studies [4-7].

One would naturally ask what the power constrains are to achieve such incredible volumes of digital content. Andrae and Edler from Huawei Technologies Sweden recently published an estimate of the global electricity usage that can be ascribed to consumer devices, communication networks and data centers between 2010 and 2030 [8]. Their estimates showed that Communication Technologies could use as much as 51% of global electricity capacity by 2030.

Here we estimate the energy and power needs to sustain the annual production of information assuming an annual growth of f % per year. Currently the energy required to write a bit of information, regardless on the data storage technology used, is much higher than the minimum predicted energy, $Q_{bit} = k_B T \cdot \ln(2) \approx 18$ meV at room temperature (T = 300K). Let us assume that our future technological progress will allow writing digital information with maximum efficiency. In this case, the total energy necessary to create all the digital information in a given $n^{th}$ year, assuming f % year-on-year growth is given by:

$$Q_{\inf o}(n^{th}) = N_b \cdot k_B T \cdot \ln(2) \cdot (f+1)^n \qquad (2)$$

The total planetary power requirement to sustain the digital information production is obtained dividing relation (2) by the number of second in a year, $t \approx 3.154 \times 10^7$.

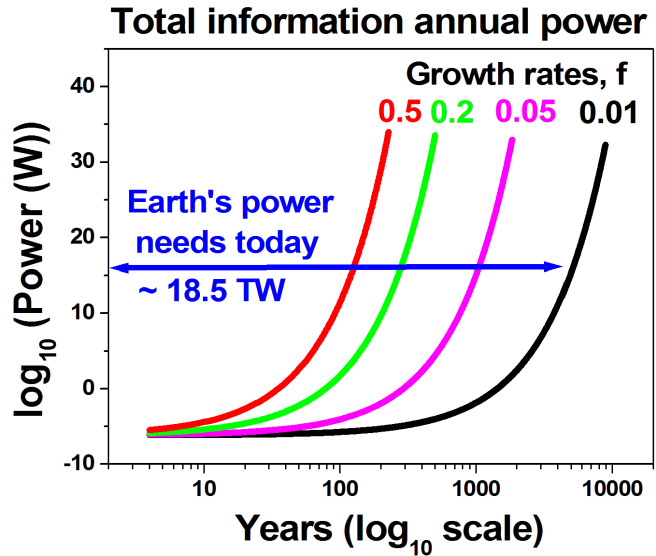

**Figure 2.** Increase in the annual power requirement to create digital information, assuming annual digital content growth rates of 1%, 5%, 20% and 50%. This is compared with toady's planetary power consumption of 18.5TW.

Figure 2 shows the annual power versus number of years in logarithmic scale, for f = 1%, 5%, 20% and 50%, respectively. The total power requirement today to power all industries, transportation and domestic energy needs on Earth is around $18.5 \times 10^{12}$ W = 18.5TW [9], i.e. in logarithmic value this is $\log_{10}(18.5 \times 10^{12}) = 13.27$.

As could be seen from figure 2, for 1% growth rate, after ~4500 years the creation of digital content will take up the equivalent of all today's planetary power requirements. Similarly, for 5%, 20% and 50% growth rates, this will occur after ~918, ~246 and ~110 years from now, respectively. It is worth to remind that these estimates have assumed production of digital content at the maximum efficiency, which is certainly not the case yet. Hence, it appears that the current growth rate is unsustainable and digital information production will be limited in future by the planetary power constrains. This limitation could be formulated in terms of entropy, as introduced by Deutscher in his book, The Entropy Crisis, in which he argues that the energy crisis is in fact an entropy crisis, because the entropy increase in the biosphere requires energy [10]. Since information production actually increases the entropy of a system, by extrapolation, producing digital information also increases the entropy of the biosphere. Interestingly, this increase in the information entropy of the biosphere could be used in reverse, to harvest energy from entropy as previously proposed [11].

**4. Information mass estimation**

We now examine this issue considering deeper physical implications. In 2019, the mass-energy-information equivalence principle was formulated [12], stating that information is physical, and it transcends into mass or energy depending on its state. Although this principle still awaits an experimental verification, assuming it is correct, it opens up interesting possibilities with wide ranging implications for computing technologies, physics and cosmology. The mass-energy-information equivalence principle explains the mechanism by which a classical digital bit of information at equilibrium stores data without energy dissipation, requiring the bit to acquire a mass equal to $m_{bit}$, while it stores information. Essentially a bit of information could be seen as an abstract information particle, with no charge, no spin and rest mass, $m_{bit} = (k_B T \cdot \ln(2))/c^2$, where $k_B$ is the Boltzmann constant, T is the temperature at which the information is stored and c is the speed of light. In fact, it was proposed that "information" is not only the fifth form of matter along solid, liquid, gas and plasma, but also possibly the dominant form of matter in the Universe [12]. Although this principle was first formulated in 2019, these ideas are not new. The legendary physicist, John Archibald Wheeler, considered the Universe made up of three parts: Particles, Fields and Information. In fact, Wheeler proposed reformulating the whole Physics in terms of the information theory. He summarized his ideas in a paper that he delivered at the Santa Fe Institute in 1989 [13], in which he postulated that the Universe emanates from the information inherent within it and he coined the phrase "It from bit." Moreover, other scientists also estimated independently the mass of a bit of information [14-18], and the informational content of the universe [19] has been the central piece to new theories and ideas including Black Holes and the holographic principle [20], the computational universe [21], the emergent gravity [22,23] and the quantized inertia [24,25].

Here we explore the implications of the mass-energy-information equivalence principle, in the context of the current digital information revolution.

Using the mass-energy-information equivalence principle, the rest mass of a digital bit of information at room temperature is $m_{bit} = 3.19 \times 10^{-38}$ Kg. We can now estimate how much information mass we are creating / converting on Earth at present, every year, as the product $N_b \times m_{bit}$. The total calculated mass of all the information we produce yearly on Earth at present is $23.3 \times 10^{-17}$ Kg. This is extremely insignificant and impossible to notice. For comparison, this mass is 1000 billion times smaller than the mass of single grain of rice, or about the mass of one E.coli bacteria [26]. It will take longer than the age of the Universe to produce 1Kg of information mass. However, the production of digital information is rapidly increasing every year and the objective of this work is to estimate the total information mass after a number of n years. Let us assume again that f % is the annual growth factor of digital content creation on Earth. This allows the estimation of the total information mass accumulated on the planet, after n years of f % growth as:

$$M_{\inf o}(n) = N_b \cdot \frac{k_B T \cdot \ln(2)}{f \cdot c^2} \cdot \left((f+1)^{n+1} - 1\right) \qquad (3)$$

Assuming a conservative annual growth of digital content creation of 1%, using (3) we estimate that it will take around ~3150 years to produce the first cumulative 1Kg of digital information mass on the planet and it will take ~8800 years to convert half of the planet's mass into digital information mass. When we input larger growth rates of 5%, 20% and 50%, respectively, these numbers become extreme. The data is represented in figure 3 in logarithmic scales. At 5% growth rate of digital content production, the first 1Kg of information mass occurs after ~675 years from now, and the half planetary mass is reached after ~1830 years. Similarly, for 20% and 50% growth rates, the numbers are ~ 188 years, ~50 years, respectively for 1Kg of information mass, and ~ 495 years, ~ 225 years for half Earth's mass, respectively. Essentially, in the extreme case scenario when our digital information production growth is sustained at 50% per year, by the year 2070 we will have 1Kg of digital bits content on the planet stored on all the traditional and cloud data storage centres, endpoints such as PCs, smart-phones, and Internet of Things (IoT) devices. Similarly, at 50% growth per year, by the year 2245, half of the planet's mass will be made up of digital bits.

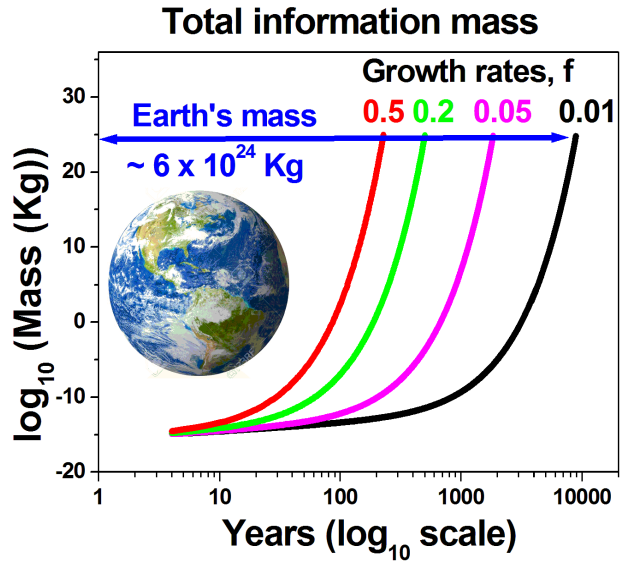

**Figure 3.** Increase in the cumulative total digital information mass created from present to n years in future, assuming annual digital content growth rates of 1%, 5%, 20% and 50%. This is compared with Earth's mass ~ $6 \times 10^{24}$ Kg.

## 5. Conclusion

In conclusion, we established that the incredible growth of digital information production would reach a singularity point when there are more digital bits created than atoms on the planet. At the same time, the digital information production alone will consume most of the planetary power capacity, leading to ethical and environmental concerns already recognized by Floridi, who introduced the concept of "infosphere" and considered challenges posed by our digital information society [27]. These issues are valid regardless of the future developments in data storage technologies. In terms of digital data, the mass-energy-information equivalence principle formulated in 2019 has not been yet verified experimentally, but assuming this is correct, then in not a very distant future, most of the planet's mass will be made up of bits of information. Applying the law of conservation in conjunction with the mass-energy-information equivalence principle, it means that the mass of the planet is unchanged over time. However, our technological progress inverts radically the distribution of the Earth's matter from predominantly ordinary matter, to the fifth form of digital information matter. One could say that we are literally changing the planet bit by bit. In this context, assuming the planetary power limitations are solved, we could envisage a future World mostly computer simulated and dominated by digital bits and computer code.

**Data Availability**

The numerical data associated with this work are freely available from the author. All the data will also be freely accessible from the University of Portsmouth repository.

**Acknowledgments**

The author acknowledges the financial support received to undertake this research from the School of Mathematics and Physics, University of Portsmouth. The author also acknowledges the valuable discussions and feedback received on this manuscript from Prof. Andrew Osbaldestin and Dr. Michal Belusky.